\documentclass{emulateapj}
\begin{document}

\title{The fate of spiral galaxies in clusters: the star formation history of the anemic Virgo cluster galaxy NGC~4569.}
\author{
A. Boselli\altaffilmark{1}, S. Boissier\altaffilmark{1,2}, L. Cortese\altaffilmark{1,6},  
A. Gil de Paz\altaffilmark{2,3}, M. Seibert\altaffilmark{4},
B. F. Madore\altaffilmark{2,5}, V. Buat\altaffilmark{1},
%, T. Barlow\altaffilmark{4}, L. Bianchi\altaffilmark{5},
%Y.-I. Byun\altaffilmark{5}, J. Donas\altaffilmark{1},
%K. Forster\altaffilmark{3}, P. G. Friedman\altaffilmark{3},
%T. M. Heckman\altaffilmark{6}, 
%P. Jelinsky\altaffilmark{7},
%Y.-W. Lee\altaffilmark{5}, 
%R. Malina\altaffilmark{1},
D. C. Martin\altaffilmark{4}
%, B. Milliard\altaffilmark{1},
%P. Morrissey\altaffilmark{3}, S. Neff\altaffilmark{9},
%R. M. Rich\altaffilmark{10}, D. Schiminovich\altaffilmark{3},
%O. Siegmund\altaffilmark{7}, 
%T. Small\altaffilmark{3}, 
%A. S. Szalay\altaffilmark{6}, 
%B. Welsh\altaffilmark{7}, T. K. Wyder\altaffilmark{3}
}
\altaffiltext{1}{Laboratoire d'Astrophysique de Marseille, BP8, Traverse du Siphon, F-13376 Marseille, France}
%\altaffiltext{2}{Universit\`a degli Studi di Milano - Bicocca, P.zza della Scienza 3, 20126 Milano, Italy}
\altaffiltext{2}{Observatories of the Carnegie Institution of Washington,
813 Santa Barbara St., Pasadena, CA 91101}
\altaffiltext{3}{Departamento de Astrof\'{\i}sica y CC. de la Atm\'osfera,
Universidad Complutense de Madrid,
Avda. de la Complutense, s/n,
E-28040 Madrid, Spain}
\altaffiltext{4}{California Institute of Technology, MC 405-47, 1200 East
California Boulevard, Pasadena, CA 91125}
%\altaffiltext{4}{Center for Astrophysical Sciences, The Johns Hopkins
%University, 3400 N. Charles St., Baltimore, MD 21218}
%\altaffiltext{5}{Center for Space Astrophysics, Yonsei University, Seoul
%120-749, Korea}
%\altaffiltext{6}{Department of Physics and Astronomy, The Johns Hopkins
%University, Homewood Campus, Baltimore, MD 21218}
%\altaffiltext{7}{Space Sciences Laboratory, University of California at
%Berkeley, 601 Campbell Hall, Berkeley, CA 94720}
\altaffiltext{5}{NASA/IPAC Extragalactic Database, California Institute
of Technology, Mail Code 100-22, 770 S. Wilson Ave., Pasadena, CA 91125}
%\altaffiltext{9}{Laboratory for Astronomy and Solar Physics, NASA Goddard
%Space Flight Center, Greenbelt, MD 20771}
%\altaffiltext{10}{Department of Physics and Astronomy, University of
%California, Los Angeles, CA 90095}
%\altaffiltext{11}{Oxford University, Astrophysics, Oxford OX1 3RH, United Kingdom}
\altaffiltext{6}{School of Physics and Astronomy, Cardiff University, 5, The Parade, Cardiff CF24 3YB, UK}

\begin{abstract}
We present a new method for studying the star formation history of
late-type, cluster galaxies undergoing gas starvation or a
ram-pressure stripping event by combining bidimensional multifrequency
observations with multi-zones models of galactic chemical and spectrophotometric
evolution. This method is applied to the Virgo cluster anemic galaxy
NGC~4569. We extract radial profiles from recently obtained UV GALEX
images at 1530 and 2310 \AA, from visible and near-IR narrow
(H$\alpha$) and broad band images at different wavelengths ($u$, B,
$g$, V, $r, i, z$, J, H, K), from Spitzer IRAC and MIPS images 
and from atomic and molecular gas
maps. The model in the absence of interaction (characterized by 
its rotation velocity and spin parameter) is constrained by the
unperturbed H band light profile and by the H$\alpha$ rotation
curve. We can reconstruct the observed total-gas radial-density
profile and the light surface-brightness profiles at all wavelengths
in a ram-pressure stripping scenario by making simple assumptions
about the gas removal process and the orbit of NGC~4569 inside the
cluster. The observed profiles cannot be reproduced by simply stopping
gas infall, thus mimicing starvation. Gas removal is required, which
is more efficient in the outer disk, inducing a radial quenching in
the star formation activity, as observed and reproduced by the
model. This observational result, consistent with theoretical
predictions that a galaxy-cluster IGM interaction is able to modify
structural disk parameters without gravitational perturbations, is
discussed in the framework of the origin of lenticulars in clusters.

\end{abstract}


\section{Introduction}

\setcounter{footnote}{0} Perturbations induced by the harsh cluster
environment makes the evolution of cluster galaxies different than
that of their counterparts in the field. Besides the well known
morphology segregation effect (Dressler 1980; Whitmore et al. 1993),
there is also evidence clearly indicating that the cluster late-type
galaxy population is systematically different from the field: cluster
spirals are generally depleted in their total HI content and are less
active in forming stars than their isolated counterparts. This is most
recently and extensively reviewed in Boselli \& Gavazzi (2006). The
nature of the perturbing phenomenon has not yet been unambiguously
identified. Several physical processes have been proposed to explain
gas removal in clusters. Some of them are related to the dynamical
interactions of cluster galaxies with the hot ($T_{IGM}$ $\sim$
10$^7$ K) and dense ($\rho_{IGM}$ $\sim$ 10$^{-3}$ atom cm$^{-3}$)
intergalactic medium (IGM) (ram-pressure, Gunn \& Gott 1972; viscous
stripping, Nulsen 1982; thermal evaporation, Cowie \& Songaila
1977). Others are due to the gravitational interactions with nearby
companions (Merritt 1983) or with the wider potential of the cluster (Byrd
\& Valtonen 1990).
%The lack of a sufficient gas reservoir needed to
%feed star formation is then at the origin of the decrease of the
%activity observed in cluster galaxies.
%\\
%All these processes, however, {\textbf are active only nearby the core of the
%clusters, where the IGM gas and galaxy densities, the gas temperature and the
%cluster potential are at their maximum.
%NOT TRUE FOR LOW VELOCITY INTERACTIONS! 
%YOU MUST SPECIFY IS LOW OR HIGH VELOCITY INTERACTIONS OR BOTH.} 
%Recent, complete spectroscopic surveys of the nearby universe such as the SDSS or the
%2dF, able to trace the radial variation of the star formation activity
%up to large cluster-centric distances, have shown that the activity of
%late-type galaxies begins to decrease at $\sim$ 1-2 virial radii from
%the cluster center (Gomez et al. 2003; Lewis et al. 2002, Tanaka et
%al. 2004, Nichol 2004), scale-lengths comparable to those observed in
%the HI (Gavazzi et al. 2005,2006a). Since these distances (1-2 virial
%radii) are significantly larger (a factor of 5-10) than those where
%galaxy-IGM or galaxy-galaxy interactions {\textbf WHAT KIND OF TIDAL INTERACTION} 
%are expected to be at place ($\sim$ inside a
%core radius), other processes more active at the periphery of clusters
%such as galaxy harassment (Moore et al. 1996) {\textbf I'M NOT SURE YOU ARE RIGHT SAYING 
%THAT HARASSMENT IS MORE ACTIVE AT THE CLUSTER PERIFERY}, starvation (Balogh et
%al. 2000) or pre-processing (Fujita 2004) have been proposed to
%explain the radial decrease of the star formation activity with the
%cluster-centric distance.
%\\
The detailed analysis of several well known resolved galaxies in the
Virgo and in the Coma clusters, generally favor a galaxy-IGM
interaction scenario (Vollmer et al. 1999, 2000, 2001a, 2004b; Kenney et al. 2004;
Yoshida et al. 2004) in these instance. Using complete, multifrequency datasets combined
with model predictions several studies have unambigously shown that
ram-pressure stripping occurs even at the periphery of
clusters (e.g. CGCG 97-73 and 97-79 in A1367, Gavazzi et al. 2001), 
and that galaxies recently ($\leq$ 500 Myr) stripped by
ram-pressure populate nearby clusters well outside the cluster core
(Boselli \& Gavazzi 2006). 
\\
%This apparent discrepancy {\textbf WHICH DISCREPANCY, HAI APPENA DETTO CHE GAV\&BOS HANNO 
%DIMOSTRATO CHE LA RAM PRESSURE E' ATTIVA ANCHE FUORI DAL CLUSTER CORE} 
%poses major limits to the understanding of the evolution of galaxies in
%different environments, i.e. 
Ram-pressure stripping has also been
invoked to explain the origin of the lenticular
galaxy population inhabiting rich clusters. 
One early suggestion was 
that, as a consequence of gas removal through ram-pressure stripping, cluster
galaxies first become ``anemic'' and then passively evolved into
lenticulars (van den Bergh 1976). Statistical considerations, as well
as structural, kinematical and spectroscopic properties of
lenticulars, however, do not appear to confirm this simple scenario. As
discussed in Dressler (1980, 2004) the morphological type variation in clusters is 
too weak a function of the local galaxy density. Furthermore, the disks of
lenticulars have, on average, surface brightnesses and bulge-to-disk
ratios that are significantly higher than early-type spirals (Dressler 1980;
Christlein \& Zabuldoff 2004; Boselli \& Gavazzi 2006) making the
formation of lenticulars through gas sweeping in spirals very
unlikely.  The larger scatter and a small zero-point offset in the
Tully-Fisher relation observed in the Virgo and Coma cluster lenticulars
also indicate that S0s can hardly be formed by simple gas removal from
healthy spirals (Hinz et al. 2003).  Furthermore cluster lenticulars
generally have spectroscopic signatures of recent, post-starburst 
activity (Poggianti et al. 2004) difficult to trigger in a simple
ram-pressure stripping scenario.\\ 

The origin of lenticulars is still an open problem and
limit our understanding of the evolution of galaxies in
different environments. This riddle can be solved only by first realistically 
quantifying the physical
effects of ram-pressure stripping by combining model predictions with
multifrequency observations of representative samples.
\\
We have been collecting multi-frequency data for a large sample of
late-type galaxies in nearby clusters and in the field in order to
undertake the comparative statistical analysis of any systematic
differences between cluster and field objects. The most important results obtained from
the analysis done so far are reported in Boselli \& Gavazzi (2006).
Combined with multi-zone models for the chemical and spectrophotometic evolution of
galaxies (Boissier \& Prantzos 2000), this unique database is helping us
to understand the evolution of cluster spirals. \\ 
The aim of the present paper is to give a complete description of the
multi-zone spectrophotometric models of galaxy evolution used to study
the effects of the environment on cluster galaxies. 
%These are
%essentially those of Boissier \& Prantzos (2000) modified to take care
%of a time-dependent gas stripping event as predicted by the dynamical
%models of Vollmer et al. (2001b) (ram pressure scenario) or
%by simply suppressing gas infall (starvation scenario). 

As a first application we present a
detailed study of the radial profiles of the Virgo cluster galaxy
NGC~4569 (M90).  NGC~4569, the prototype anemic galaxy as defined by
van den Bergh (1976), is extremely deficient in HI, having only about 
one tenth of the atomic gas of a comparable field galaxy of similar
type and dimensions. This galaxy has a truncated H$\alpha$ and HI
radial-density profile (at a radius of $\sim$ 5 kpc; see Fig.~\ref{images} and \ref{FIGmodelREF}) as firstly
noticed by Koopmann \& Kenney (2004a,b) and Cayatte et al$.$ (1994),
bearing witness to a recent interaction with the cluster environment. NGC~4569
is located close ($\sim$ 2 degrees, $\sim$ 240 kpc for a virial radius of 1.7 Mpc) to the cluster center. Being one of
the largest galaxies ($\sim$ 10 arcmin, $\sim$ 50 kpc) in the Virgo cluster, NGC~4569
is the ideal candidate for our study since it is  spatially
resolved at all the wavelengths considered here. NGC 4569 has also been
selected because it has been previously  studied using dynamical models by Vollmer et
al. (2004a). A direct comparison of the results obtained using totally
independent techniques (in particular, dating the interaction)  is
thus possible. 
We just remind that multifrequency and kinematical
observation of the inner part of NGC 4569 done by Jogee et al. (2005)
seem to indicate that the nucleus and the inner stellar bar of the
galaxy might have been triggered by a recent tidal interaction. There are however no observational evidences 
indicating that the HI disk has been truncated by any gravitational interaction with nearby companions.
%Our
%assumption of a ram-pressure stripping event at the origin of the
%radial truncation of the gas and stellar disk of NGC 4569 is not in
%contradiction with the Jogee et al. result for two reasons: as
%discussed in Boselli \& Gavazzi (2006) tidal interactions might induce
%some gas infall in the nuclei of cluster galaxies, but can hardly
%remove the outer disk gas {\textbf E' VERO CHE LA TIDAL NON TRONCA MA POTREBBE 
%MODIFICARE LA NORMALIZZAZIONE DEL PROFILO PORTANDO GAS AL CENTRO. IO LASCEREI SOLO 
%LA SECONDA MOTIVAZIONE. INOLTRE NON È ANCORA CHIARO SE NGC4569 HA AVUTO TIDAL, INFATTI 
%COME DICE VOLLMER, POTREBBE ESSERE ANCHE SOLO UNO STARBURST DOVUTO A SECULAR EVOLUTION ED ALLA 
%PRODUZIONE DELLA BARRA AL CENTRO}. 
Since our models are adapted for disk galaxies, we will however limit our study to the disk component 
excluding the central nucleus.\\ 
The combination of the
multi-frequency 2-D data with our spectrophotometric models allow us
to study, for the first time, the radial evolution of the different
stellar populations in this prototype, gas-stripped cluster galaxy
with the aim of understanding whether its structural properties can
evolve, even in principle,  into those of a typical cluster lenticular (S0) galaxy.  A more
complete analysis of a statistically significant sample of cluster
galaxies will be presented in a future communication.

\section{Data}

%%%SAM (on parle des modeles plus loin)
%The 2-D chemo-spectrophotometric models of galaxy evolution of
%Boissier \& Prantzos (2000) reproduce the radial profiles of the
%different stellar components, of the total (HI + H$_2$) gas content,
%of the metallicity as well as the rotation curve of galaxies. 

\begin{figure*}
\epsscale{1.0} \includegraphics[width=15cm,angle=0]{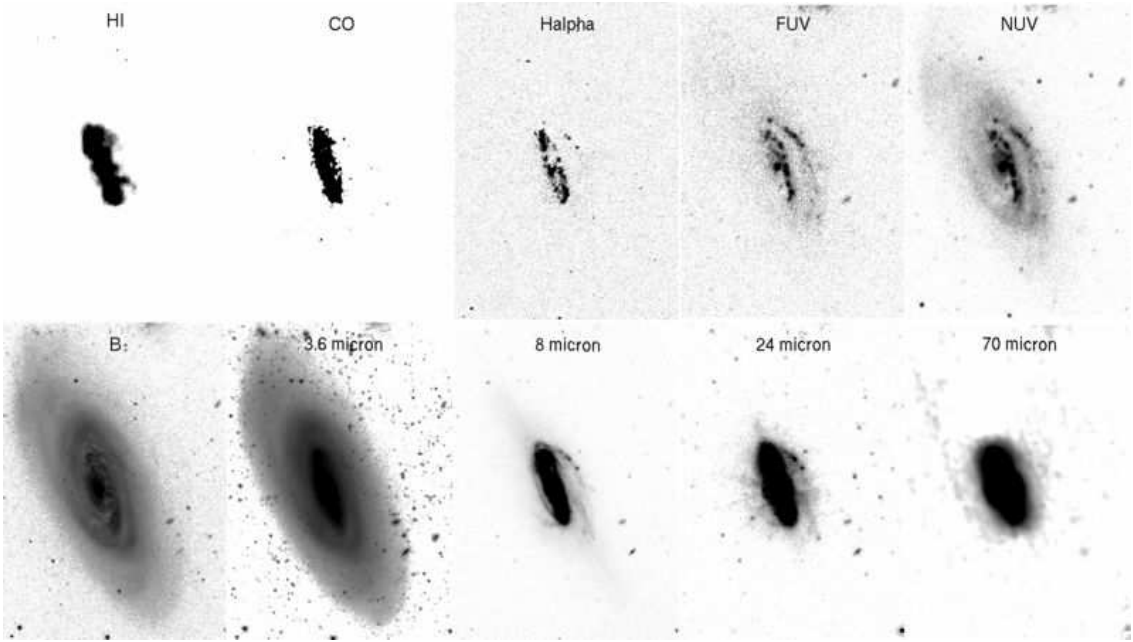}
\small{\caption{The multifrequency images of NGC 4569 at their intrinsic resolution: from top left
to bottom right, HI, CO, H$\alpha$, FUV, NUV, B, 3.6 $\mu$m, 8$\mu$m, 24 $\mu$m and 70 $\mu$m all on the same scale.
\label{images}}}
\end{figure*}

The large amount of spectrophotometric data available for NGC~4569,
collected in the GOLDMine database (Gavazzi et al$.$ 2003), allow us
to reconstruct its radial profile at 22 different wavelengths: from the
new GALEX UV bands (at FUV=1530 and NUV=2310 \AA, IR1.0 data release recently 
published in Gil de Paz et al. 2006), 
to the visible B and V (Boselli et al$.$ 2003), Sloan $u,g,r,i,z$ 
(Abazajian et al$.$ 2005), near-IR J, H
and K bands (Boselli et al$.$ 1997; 2MASS Jarrett et al$.$
2003), mid-IR 3.6, 4.5, 5.8 and 8 $\mu$m IRAC and 
far-IR 24, 70 and 160 $\mu$m MIPS bands recently obtained by
Spitzer in the framework of the SINGS project (Kennicutt et al. 2003). 
An accurate description of the IRAC and MIPS Spitzer data reduction procedures
is given in Dale et al. (2005).
H$\alpha$+[NII] narrow band imaging, used to trace the recent
star formation activity (e.g. Boselli et al$.$ 2001), is available
from Boselli \& Gavazzi (2002). Some of the multifrequency images of NGC 4569 
are shown in Fig.~\ref{images}. The accuracy of photometry from the
imaging data is, on average about 10\%.  HI profiles are from
Cayatte et al$.$ (1994), while H$_2$ profiles, determined from CO data
using a luminosity-dependent CO-to-H$_2$ conversion factor (from
Boselli et al$.$ 2002, applied to the whole profile with no radial
variation as unfortunately no metallicity gradient information is
available for NGC~4569)
%(SAM< THIS IS A CONSTANT VALUE OR A  RADIAL DEPENDENT VALUE?)
are taken from the BIMA survey of Helfer et al$.$ (2003) for the inner
disk, and from Kenney \& Young (1988) for the outer disk. The total
gas profile is the sum of the HI and H$_2$ gas profiles multiplied by
a factor of 1.4 to take into account the He contribution.  The galaxy
rotation curve has been taken from Rubin et al$.$
(1999).\\

\begin{figure}
\epsscale{1.0} \includegraphics[width=7cm,angle=0]{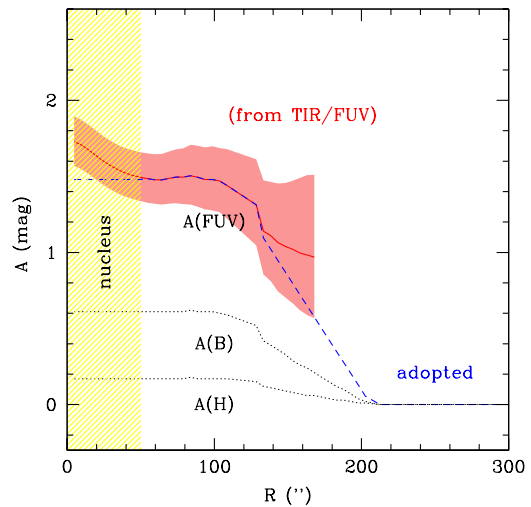}
\small{\caption{ The radial variation of the A(FUV) extinction profile (red solid line)
and its uncertainty (red shaded area) as determined from the total far-IR (TIR) to FUV flux ratio.
The adopted A(FUV) extinction correction, determined excluding the contribution of the active nucleus
and extrapolating the profile to zero values in the outer disk, is shown by the blue dashed line. 
The black dotted lines give the extinction profiles in the B and H band.
\label{extinction}}}
\end{figure}

The radial surface-density profiles have been constructed by integrating the available
images within elliptical, concentric annuli. The ellipticity and
position angles have been determined and then fixed using the deepest
B-band image following the procedure of Gavazzi et al$.$ (2000)\footnote{ 
To avoid any possible contamination by the NW arm, whose
kinematical properties indicate that it is not assoicated to the
stellar disk but rather formed during the galaxy-cluster IGM
interaction, the arm was masked in the construction of the radial
profiles. If included, its contribution would be perceptible only in
the FUV filter at radii $>$ 8 kpc, increasing the surface brightness
by $<$ 0.5 mag.}. 
All the observed profiles have been smoothed to the same angular resolution
as the models (which is 1~kpc, see next section).
The UV to near-IR broadband radial profiles of the galaxy
have been corrected for internal extinction using the recipe of
Boissier et al$.$ (2004) based on the radial variation of the far-IR to FUV flux ratio.
After masking the contribution of the nucleus, whose emission is contaminated by 
AGN activity (NGC 4569 is classified as a Seyfert in NED), we combined the 24, 70 and 160 $\mu$m
profiles, smoothed to the 160 $\mu$m resolution, using the receipe of Dale et al. 
(2001) to estimate the total 1-1000 $\mu$m dust emission (TIR) as extensively discussed in Cortese et al. (2006a).
The total far-IR to FUV flux ratio radial variation is then transformed into a
FUV extinction gradient (in magnitudes) using the appropriate calibration given in Cortese et al. 
(in preparation) for a galaxy with spectral properties similar to those of NGC 4569. 
We stress that, given its quiescent nature, the contribution of far-UV photons to the
dust heating of NGC 4569 is only marginal: the calibration of the far-IR to FUV flux ratio into a
FUV extinction is thus more indirect than in star forming galaxies, 
where dust is mostly heated by UV photons, making the correction highly uncertain.
The adopted far-IR to FUV flux ratio vs. A(FUV) relation calibrated on the spectral energy 
distribution of NGC 4569 allow us to take into account also the contribution to dust heating from the general 
interstellar radiation field not necessarily associated to star forming regions. 
Indeed we predict less extinction for the same TIR/FUV ratio than for star forming galaxies (Buat et al. 2005). 
The extinction in the other visible and near-IR
bands has been determined using the prescription of Boselli et al$.$
(2003). The resulting A(FUV) extinction profile, along with A(B) and A(H), 
are shown in Fig.~\ref{extinction}: given the lack of gas and dust in the outer disk,
extinctions are extrapolated to zero at large radii. As extensively discussed in the next sections, the
determination of the radially-dependent extinction corrections,
that can be quite large (i.e., more than 1~mag) in the UV bands, is thus the major source
of systematic uncertainty in the reconstruction of the corrected radial
profiles.\\ 

\begin{figure*}
\epsscale{1.0} \includegraphics[width=15cm,angle=0]{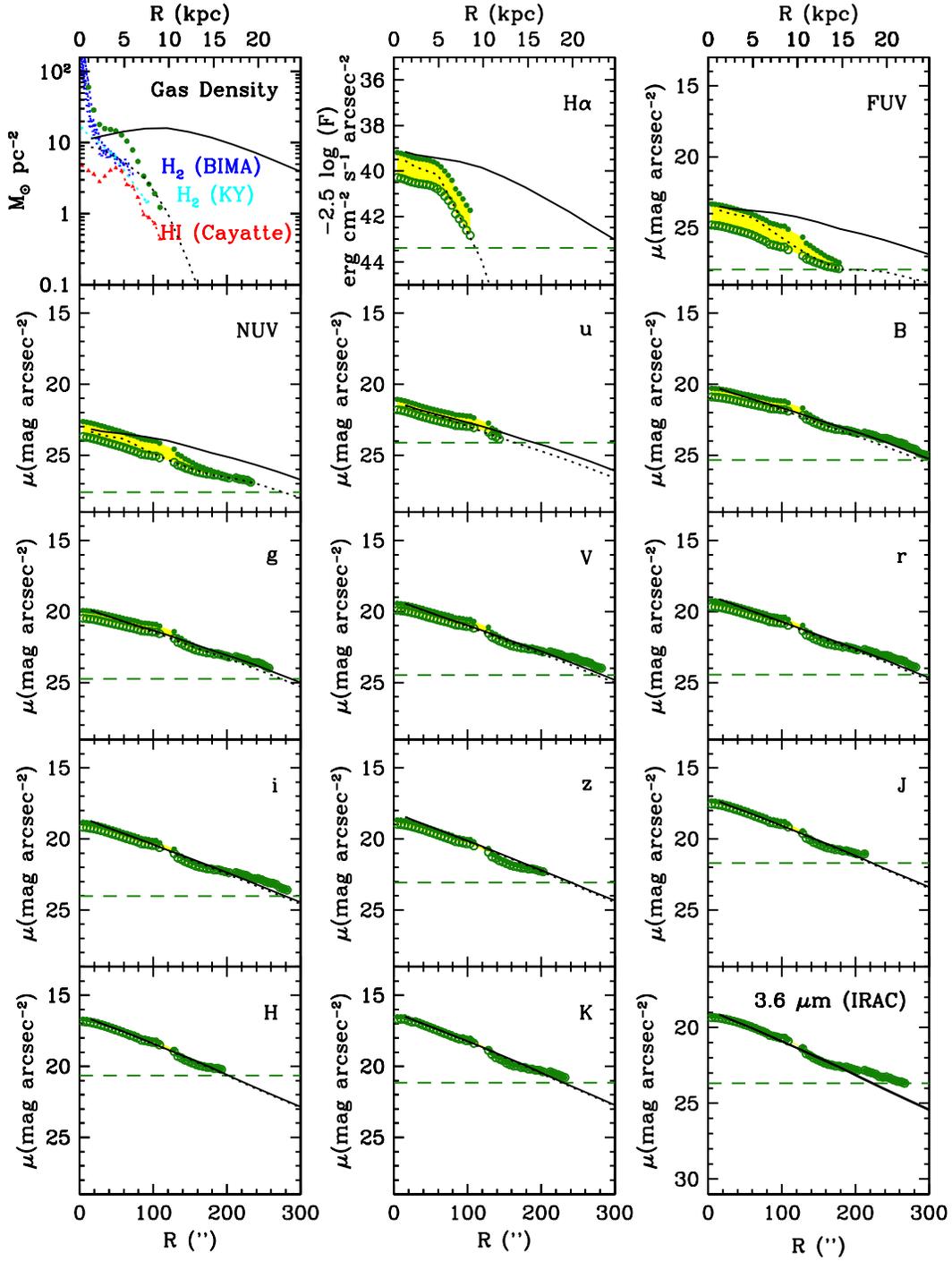}
\small{\caption{Profiles in NGC~4569. For the photometry, the open
symbols are the observed values, the filled symbols the ones corrected
for extinction. The dashed horizontal line in each panel indicates
detection limits, determined as described in Gil de Paz \& Madore (2005) 
for a S/N $\sim$ 1, where the rms is a combination of the pixel per pixel and 
large scale sky background rms. The total gas profile (top left panel: filled
circles) is computed from the sum of the HI from Cayatte et al. 1994 (filled
triangles), H$_2$ from Helfer et al$.$ 2003 (BIMA) and Kenney \& Young
1988 (KY); corrected for a Helium contribution ($\times$ 1.4). The
observations were smoothed to 1~kpc,  matching the resolution
used in the models. The solid and dotted lines are respectively the
reference model, and our best-model including ram pressure (see text
for details on the models).
\label{FIGmodelREF}}}
\end{figure*}

\begin{figure}
\epsscale{1.0} \includegraphics[width=7cm,angle=0]{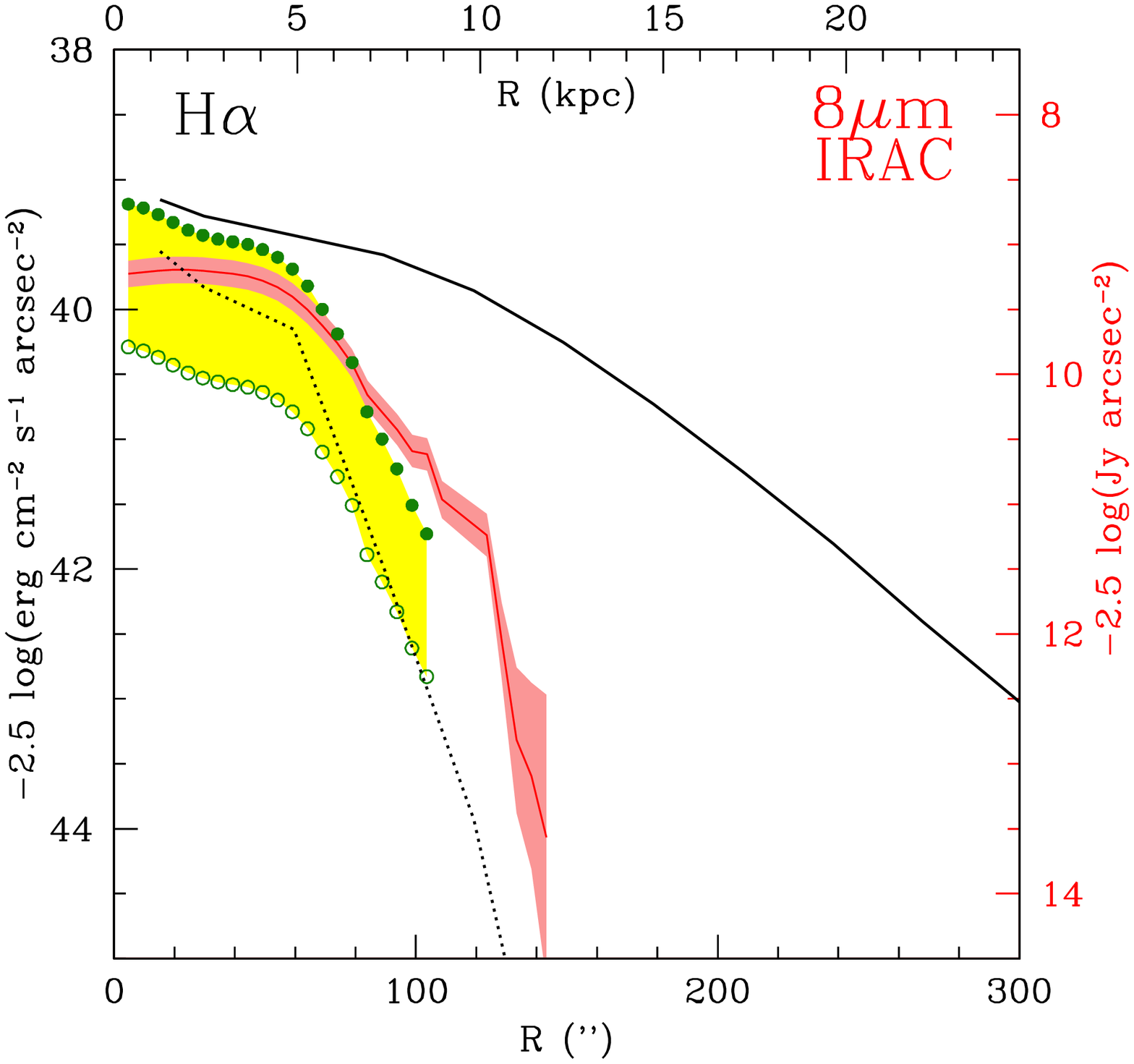}
\small{\caption{The comparison of the observed (green empty circles) and corrected (green filled circles) H$\alpha$ 
and 8 $\mu$m IRAC (red line) radial profiles. Both the H$\alpha$ and the 8 $\mu$m IRAC (red line) radial profiles are
strongly truncated if compared to the unperturbed model (black solid line), and qualitatively similar to the 
radial variation of the star formation activity predicted by the adopted perturbed model in a ram pressure 
scenario (black dotted line).
\label{prof8mic}}}
\end{figure}

H$\alpha$+[NII] narrow-band imaging has been corrected
for [NII] contamination and dust extinction (Balmer decrement) using
the integrated spectroscopy of Gavazzi et al$.$ (2004). Given the
nature of the integrated spectrum, these corrections are fixed  and
do not change with radius. We believe that such a constant
correction does not introduce systematic errors in the H$\alpha$ profile
determination given the strongly truncated morphology of the
galaxy. It is generally considered that radial effects are small for
these corrections (e.g., Martin \& Kennicutt 2001). On the other hand,
the integrated spectrum might be partly contaminated by the nuclear
emission (the galaxy is classified as a Seyfert in NED). In any case [NII]
contamination and extinction are the two major sources of
uncertainty in the determination of the H$\alpha$ radial profile.\\
In Fig.~\ref{FIGmodelREF}, we show all the gas and stellar profiles described above,
at 1 kpc resolution. For the surface photometry, open
symbols correspond to observed values, and filled symbols to
extinction-corrected values. The shaded area in between the two profiles 
graphically illustrates the magnitude of the uncertain extinction correction.
We notice that, although not univocally related to star formation (Boselli et al. 2004), the unestinguished
8 $\mu$m image (Fig.~\ref{images}) and radial profile (Fig. ~\ref{prof8mic}) of 
NGC 4569 confirm the truncated nature of the star forming disk.

\section{Models}

\subsection{The multi-zone models for the chemical and spectro-photometric unperturbed disk evolution}
To study the evolution of the disk of NGC~4569 at various radii, we
have used the multi-zone (typical resolution $\sim$1 kpc) chemo-spectrophotometric models of Boissier
\& Prantzos (2000), updated with an empirically-determined star
formation law (Boissier et al$.$ 2003) relating the star formation
rate to the total-gas surface densities ($\Sigma_{SFR}$, $\Sigma_{gas}$):
\begin{equation}
\Sigma_{SFR}= \alpha \Sigma_{gas}^{1.48} V(R)/R
\end{equation}
where $V(R)$ is the rotation velocity at radius $R$. 
%It is in fact
%very similar to the one used in the original models of Boissier \&
%Prantzos (2000) where the index was 1.5 rather than 1.48, and the
%global efficiency was slightly higher. 
The resulting models are
extremely similar to those in Boissier \& Prantzos (2000) and show
the same global trends. Not only do we consider the star formation law as
fixed but we also keep the same mass accretion (infall) histories as
in Boissier \& Prantzos (2000), based on the assumption that before the
interaction with the cluster, NGC4569 was a ``normal'' spiral.  The
only remaining free parameters in this grid of models are the spin
parameter, $\lambda$ and rotational velocity, $V_C$. 
The spin parameter is a dimensionless measure of the angular momentum
(defined in e.g. Mo, Mao \& White 1998). Its value in spirals ranges
typically between $\sim$ 0.02 for relatively compact galaxies to
$\sim$ 0.09 for low surface brightness galaxies (Boissier \& Prantzos
2000). The models of Boissier \& Prantzos (2000) contains scaling
relationships (the total mass varies as $V_C^3$, the scale-length as
$\lambda \times V_C$). Star formation histories depend on the infall
timescales, which are a function of $V_C$ in these models, so that 
roughly speaking,
$V_C$ controls the stellar mass accumulated during
the history of the galaxy, and $\lambda$ its radial distribution.
\begin{figure}
\epsscale{1.0}
%\plotone{go5petit.ps}
\includegraphics[width=8cm]{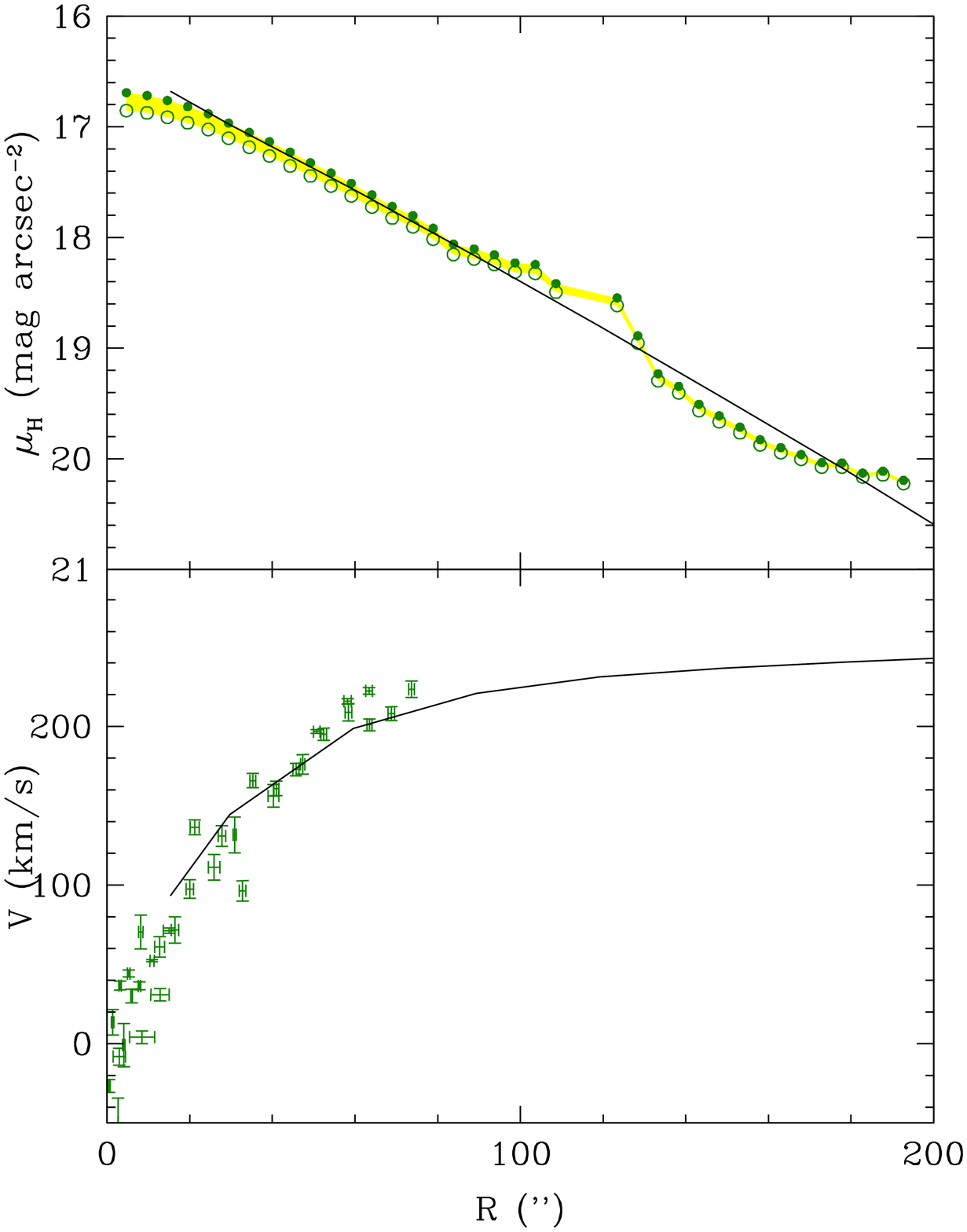}
\small{\caption{The radial profile of observed (open symbols)
and extinction-corrected (filled symbols) 
H-band surface brightness (top) and of the
rotational velocity (bottom) used to constrain the model without
interaction (represented by the black solid line). The inner 3 kpc (40 '')
are not used to constrain the model. 
\label{FIGconstraints}}}
\end{figure}
To constrain these two parameters, we use the H-band luminosity
profile (determined assuming a distance of 17 Mpc) and the rotation
curve of the galaxy, making the reasonable assumption that both of
these observables are unperturbed during the interaction\footnote{The
H-band luminosity, proportional to the total dynamical mass of the
galaxy (Gavazzi et al$.$ 1996) and tracer of the old stellar
population, and the rotational velocity of galaxies are generally not
affected by the interaction with the harsh cluster environment
(Boselli \& Gavazzi 2006).}.  Since the model does not include any
bulge or nuclear component, throughout the paper the comparison
between models and observed profiles is limited to the disk component,
excluding the inner $\sim$ 40'' (3 kpc). In order to fix the two parameters,
we proceeded as follow: for each spin considered (0.01 to 0.09 in 0.01
steps), we computed models with various rotational velocities (80 to
360, in steps of 70~km s$^{-1}$) and interpolated the models for each
spin in order to find the velocity for which the integrated H band magnitude
is equal to the observed one, after 13.5~Gyr of evolution (considered as
the present epoch). We then compared the model profile in the H band
for each spin (with the velocity determined as described above) to the
observed profile. The best agreement was clearly obtained with
$\lambda$=0.04 and the associated $V_C$=270 km s$^{-1}$. The disk is
more concentrated (extended) than observed for smaller (larger) spin
parameters. The resulting rise of the rotation curve for this model is
consistent with the observed one 
%, and the H band surface brightness is well
%reproducing the exponential decrease with radius 
(see Fig.~\ref{FIGconstraints} for a comparison of this model and 
the constraints mentioned above). 
The model does not reproduce small scales variations probably due to
structures such as spiral arms. Given the agreement, 
this model will now be considered as  the
reference model for the unperturbed case.
The models of Boissier \&
Prantzos (2000) provide the luminosity profiles in all bands as well
as the total gas profile. They do not compute the nebular emission,
but we estimated the H$\alpha$ emission here by using the number of
ionizing photons predicted by Version 5 of STARBURST 99 (Vazquez \&
Leitherer 2005) for a single generation of stars distributed on the
Kroupa et al. (1993) initial mass function (as used in our models),
convolving it with our star formation history, and converting the
result into our  H$\alpha$ flux following Appendix A of Gavazzi et
al. (2002b). All the profiles predicted for the reference model
(without any interaction, solid line) are compared to the observed one in
Fig.~\ref{FIGmodelREF}. The profiles at long wavelengths are in
agreement with the model, while short-wavelength observations, the star
formation tracers (UV, H$\alpha$), and the gas profiles present
truncations and/or shorter scale-lengths with respect to the
expectations of the unperturbed model.

\subsection{The starvation scenario}
In the starvation scenario (Larson et al. 1980, Balogh et al. 2000,
Treu et al. 2003), the cluster acts on large scales by removing any
extended gaseous halo surrouding the galaxy, preventing further infall
of such gas onto the disk. The galaxy then become anemic simply because it 
exhausts the gas reservoir through on-going star formation.

Infall is a necessary assumption in models of the chemical
evolution of the Milky Way to account for the G-dwarf metallicity
distribution (Tinsley 1980) and is supported by some chemo-dynamical
models (Samland et al. 1997). As the disk galaxy models were obtained
through a generalisation of the Milky Way model, infall is present in
all our models. It is a schematic but simple way to describe the
growth of any galaxy from a proto-galactic clump in the distant past to
a massive present-day galaxy. Infall time scales in the models were
chosen to reproduce the properties of present day normal
galaxies (Boissier \& Prantzos 2000, Boissier et al. 2001). A radial
variation of infall time-scales is suggested by dynamical models
(Larson 1976). A prescription for this variation was
implemented in our models of the Milky Way and of spirals, allowing to
reproduce many profiles and abundance gradients in our Galaxy
(Boissier \& Prantzos 1999, Hou et al. 2000), as well as colour and
abundance gradients in external galaxies (Prantzos \& Boissier 2000).

Stopping infall (in order to mimic starvation)
at a given time is straightforward to include in the model. We shall
call $t_s$ the elapsed time since the infall termination (look back time).
In order to affect the evolution of the galaxy, starvation
must have occured before most of the gas had been accreted onto the
disk. Regions where infall occurs early-on with respect to the
starvation epoch will barely be affected, while those where infall
occurs late in the history of the galaxy will never be built up.
Starvation is a global effect ($t_s$ has no dependence on
radius); however, the infall time-scale increases with radius (inside-out
formation of the galaxy). We can therefore expect an effect on the gas and
stellar profiles since starvation will affect the outer regions
(since these form late in an isolated galaxy model) more than
inner regions (which have already formed at earlier times).

\subsection{The ram pressure stripping scenario}

In addition to the starvation scenario we can also study 
the effect of ram-pressure gas stripping. For simplicity, we
adopt the plausible scenario of Vollmer et al$.$ (2001b)
explicitly tailored to Virgo: i.e., the galaxy being modelled 
has crossed the dense IGM only once, on an elliptical orbit. The ram
pressure exerted by the IGM on the galaxy ISM varies in time following
a gaussian profile, whose peak at ($t$=$t_{rp}$) is when the galaxy is
crossing the dense cluster core at high velocity
($t$ and $t_{rp}$ are look-back times, where the present epoch
corresponds to $t$=0). The gaussian has a
width $\Delta t$ = 9 $\times$ 10$^7$ years (see Fig$.$~3 of Vollmer et
al$.$ 2001b). We make the hypothesis that the gas is removed at a rate
that is directly proportional to the galaxy gas column density
$\Sigma_{gas}$ and inversely proportional to the potential of the
galaxy, measured by the total (baryonic) local density
$\Sigma_{potential}$ (provided by the model).  The gas-loss rate
adopted is then equal to $\epsilon$
$\frac{\Sigma_{gas}}{\Sigma{potential}}$, with the efficiency
$\epsilon$ following a gaussian having  a  maximum $\epsilon_0$ at the time
$t_{rp}$, chosen to mimic the variation of the ram pressure suggested
by Vollmer et al. (2001b). This very simple, but physically-motivated
prescription should allow us to model the gas removal from the galaxy
using only two free parameters ($t_{rp}$ and $\epsilon_0$) to age-date
and measure the magnitude of this effect.
We make the further assumption that no extra star formation is induced
during the interaction. This assumption is reasonable since both
models (Fujita 1998, Fujita \& Nagashima 1999) and observations
(Iglesias-Paramo et al. 2004) do not show any significant increase of
the star formation activity in galaxies thought to be currently 
undergoing a ram-pressure stripping event.

\section{The star formation history of NGC~4569: model predictions}

\subsection{The starvation scenario}

\begin{figure}
\centering
\epsscale{1.0} \includegraphics[angle=0,width=8cm,angle=0]{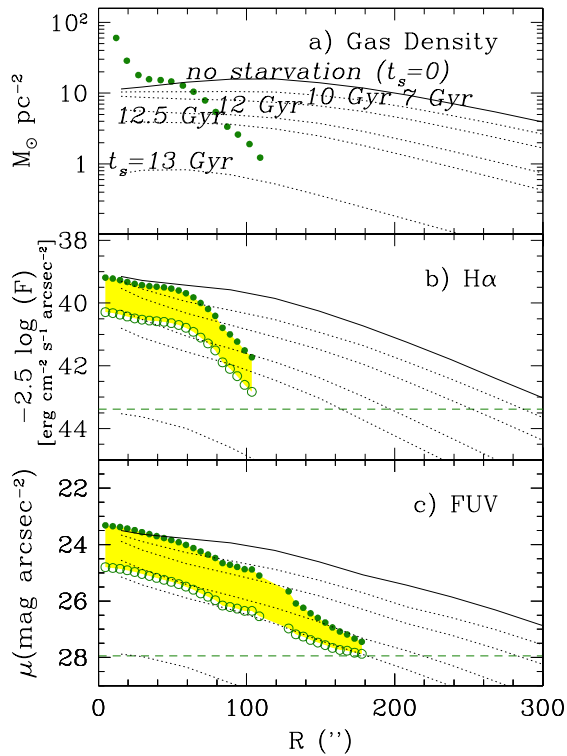}
\small{\caption{
The observed (empty circles) and corrected (filled circles) a) gas, b) H$\alpha$ and c) FUV profiles are compared to
the predictions for the reference model (solid line) and starvation
models (in which infall has been stopped for the time $t_s$, indicated
on top of each dotted curves). The dashed line correspond to our detection
limit.
\label{figstarvation}}}
\end{figure}

\begin{figure*}
\epsscale{1.0}
%\plotone{go8bNEW.ps}
\includegraphics[angle=-90,width=15cm]{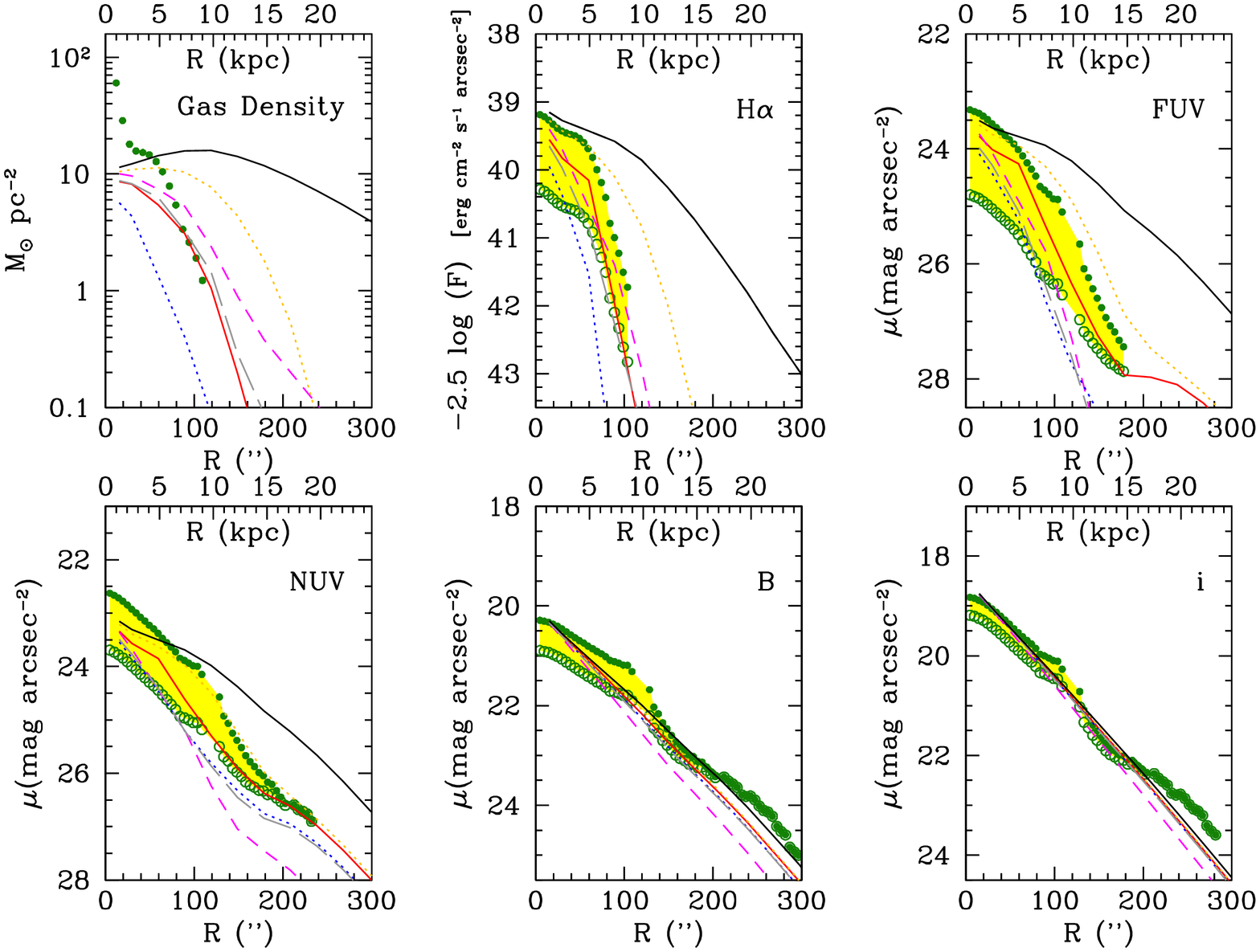} \small{\caption{The
radial profile of the observed (empty green circles) and
extinction-corrected (filled green circles) total gas, H$\alpha$, FUV
(1530 \AA), NUV (2310 \AA), B and $i$ surface brightness. The yellow
shaded area marks the range in between the observed (bottom side) and
extinction-corrected (top side) surface brightness profiles. Surface
brightnesses are compared to the model predictions without interaction
(black solid line) or with interaction (ram-pressure stripping) for several $\epsilon_0$ and
$t_{rp}$ parameters.  Equal maximum efficiency ($\epsilon_0$=1.2
M$_{\odot}$ kpc$^{-2}$ yr$^{-1}$) and different age: $t_{rp}$=100 Myr,
red continuum line (the adopted model); $t_{rp}$=300 Myr, grey long dashed
line (time suggested by the Vollmer et al. 2004a study),
$t_{rp}$=1.5 Gyr, dashed magenta line.  Equal age ($t_{rp}$=100 Myr)
and different maximum efficiency: $\epsilon_0$=3 M$_{\odot}$
kpc$^{-2}$ yr$^{-1}$, blue dotted line; $\epsilon_0$=1/3 M$_{\odot}$
kpc$^{-2}$ yr$^{-1}$, orange dotted line.
}
\label{figmodels}}
\end{figure*}

The result of starvation on the gas and star forming (H$\alpha$ and FUV) profiles,
once gas infall on the galaxy has been stopped at different epochs
($t_s$), is shown in Fig.~\ref{figstarvation}.\\
First, we note that in order to remove significant amounts of gas in
N4569, infall must have been stopped for many Gyr. This is  consistent with Balogh et al. (2000).  
The reason for this is that with $V_C$=270 km s$^{-1}$, the reference model has a higher mass
than the Milky Way. Since infall timescales in the models of Boissier
\& Prantzos (2000) are shorter for more massive galaxies, NGC4569
should have accreted most of its material early-on in its history,
thus stopping infall at a later time has no effect, as most of the gas is
already in the disk, and the galaxy is barely affected by the 
starvation process. Given the strong relationship between gas and star formation,
the same conclusion applies to the H$\alpha$ radial profile.
Secondly, the resulting profiles never have the right shape, and never predict
a sharp truncation, as is observed in all the short time-scale indicators (gas, H$\alpha$, 8 $\mu$m). 
While the models include a radial variation of
infall time-scales (inside-out formation), this trend with galactocentric distance
is not strong enough for the gas, H$\alpha$ and FUV profiles to be truncated when infall
is globally stopped at a given time.\\

The dependence of the infall time scale on the mass of the
galaxy and on the radius are the ones encoded in the models. They
would need to be dramatically changed in order to match the gas
profile observed in NGC4569 in a starvation scenario. However, if the
mass-dependence was extremely different, the models of Boissier \&
Prantzos (2000) would fail to reproduce other relations (e.g., the observed color
magnitude relationship of nearby galaxies). The radial dependence of
infall time-scales in these models is such that it provides a good match
to the color and
abundance gradients of spirals (Prantzos \& Boissier 2000). This
agreement would also be broken if we drastically changed this radial trend,
as it would be needed to obtain a truncated profile, similar to the
one observed in NGC4569.

\subsection{The ram pressure stripping scenario}
\noindent
Having such a simple prescription with only two free parameters 
(as explained in section 3.3), it is possible 
to choose simultaneously $t_{rp}$ and $\epsilon_0$
because the amount of gas left and its radial distribution depend
strongly on $\epsilon_0$ while the resulting stellar light profiles
depend mainly on $t_{rp}$ (see Fig$.$\ref{figmodels} for some examples).
\begin{figure}
\centering
\epsscale{1.0} \includegraphics[width=8cm,angle=0]{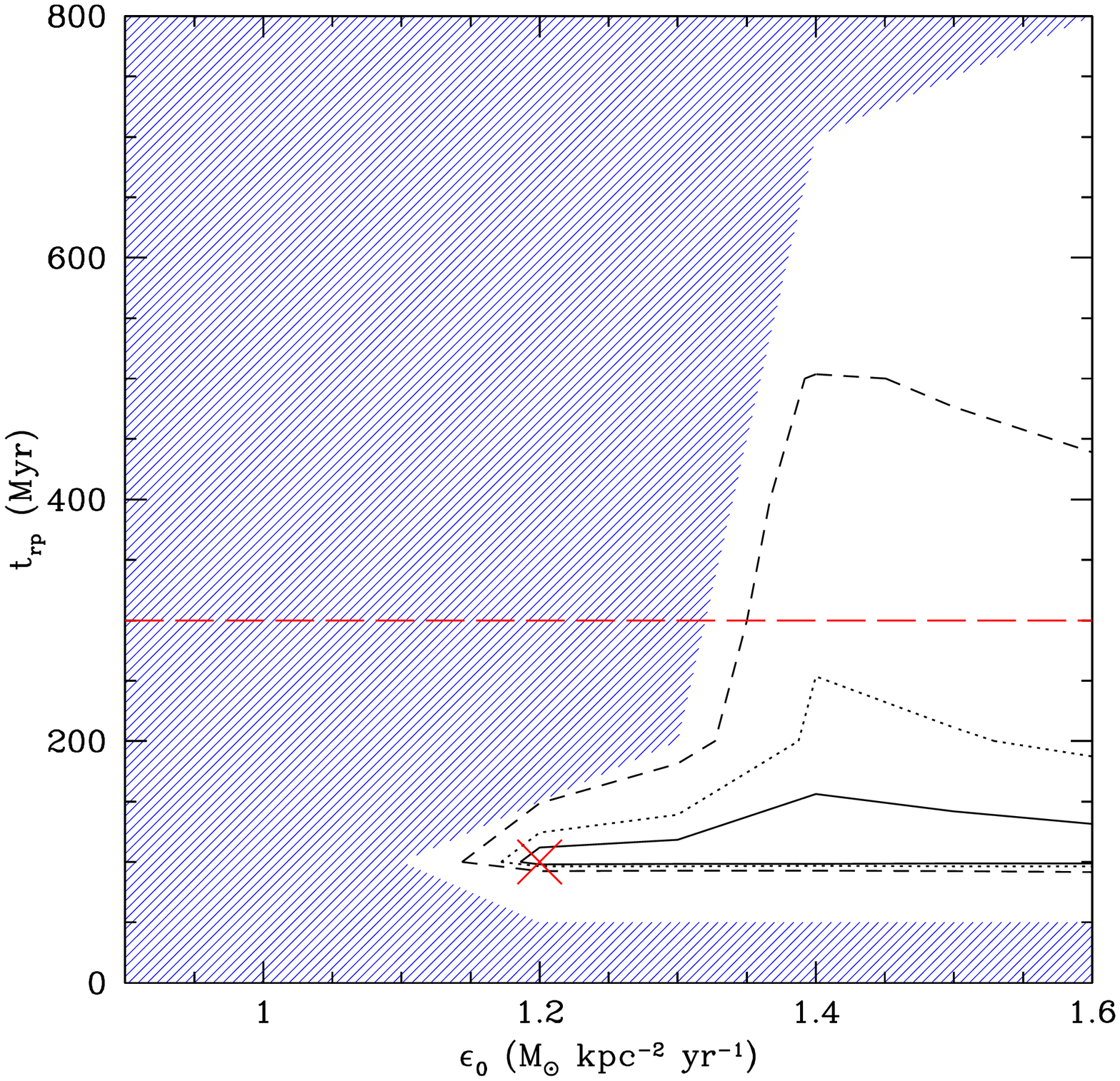}
\small{\caption{\label{figchi2}
Reduced $\chi^2$ as a function of the 
look-back time of the ram-pressure event $t_{rp}$ and efficiency
$\epsilon_0$. The contours (solid, dotted, dashed) show $\chi^2$ levels
of respectively 2,3,5 $\times$ $\chi_{min}$ (the best model gives a value
of $\chi_{min}$ $\sim$ 3.4; a ``perfect fit'' with $\chi^2 \sim$ 1 cannot be achieved
since the models will never reproduce observed small scales variations).
%{\textbf SONO VALORI ESTREMAMENTE ALTI! E' CHIQUADRO 
%RIDOTTO O NORMALE. COMUNQUE SIA QUESTI VALORI NON DICONO NIENTE BISOGNA 
%METTERE DELLE PERCENTUALI} 
The horizontal long-dashed line indicates the
time suggested by the dynamical model of Vollmer et al. (2004a).  The
shaded area indicates values for which the model surface brigtnesses
are in disagreement with observational limits (non detections at
relatively large radii).  In this case, the $\chi^2$ was arbitrarily
given large values in order to reject these solutions that would not
remove enough gas to properly reproduce the observed truncation.}}
%Reduced $\chi^2$ as a function of the look-back time of the
%ram-pressure event for a few efficiencies $\epsilon_0$ (M$_{\odot}$
%kpc$^{-2}$ yr$^{-1}$). Models were computed each 100 Myr for 0 $<$
%$t_{rp}$ $<$ 500 Myr, 200 Myr for 0.5 $<$ $t_{rp}$ $<$ 1.5 Gyr and 1 Gyr for
%1.5 $<$ $t_{rp}$ $<$ 6.5 Gyr; and each 0.2 M$_{\odot}$ kpc$^{-2}$
%yr$^{-1}$ efficiencies between 0.4 and 1.6 (only the more relevant are
%shown here). An important modification with respect to the usual
%$\chi^2$ is that its value was artificially put to 100 for any model
%predicting surface brigtnesses in disagreement with observational
%limits (non detections at relatively large radii) in order to reject
%these solutions that would not reproduce properly the observed truncation. 
%}}
\end{figure}
If the time of  cluster-core crossing  is recent 
%(i.e., $t_{rp}$ is close to the age of the galaxy, assumed to be 13.5 Gyr), 
only the youngest stellar populations (emitting in H$\alpha$, whose age is $\leq$ 4
10$^6$ yrs, or far-UV, $\leq$ 10$^8$ yrs) have had time to feel the
progressive radial suppression of the star formation activity and
we can date the suppression of gas with our model predictions.
%with the spectrophotometric radial
%profiles of cluster galaxies can thus be used to date the dynamical
%interaction with the IGM.  
We stress that this model is
principally constrained by the radial variation of the surface
brightness and color profiles and only in a minor way by their absolute
values which can be affected by zero-point uncertainties. 
\begin{figure}
\centering
\epsscale{1.0} \includegraphics[width=8cm,angle=0]{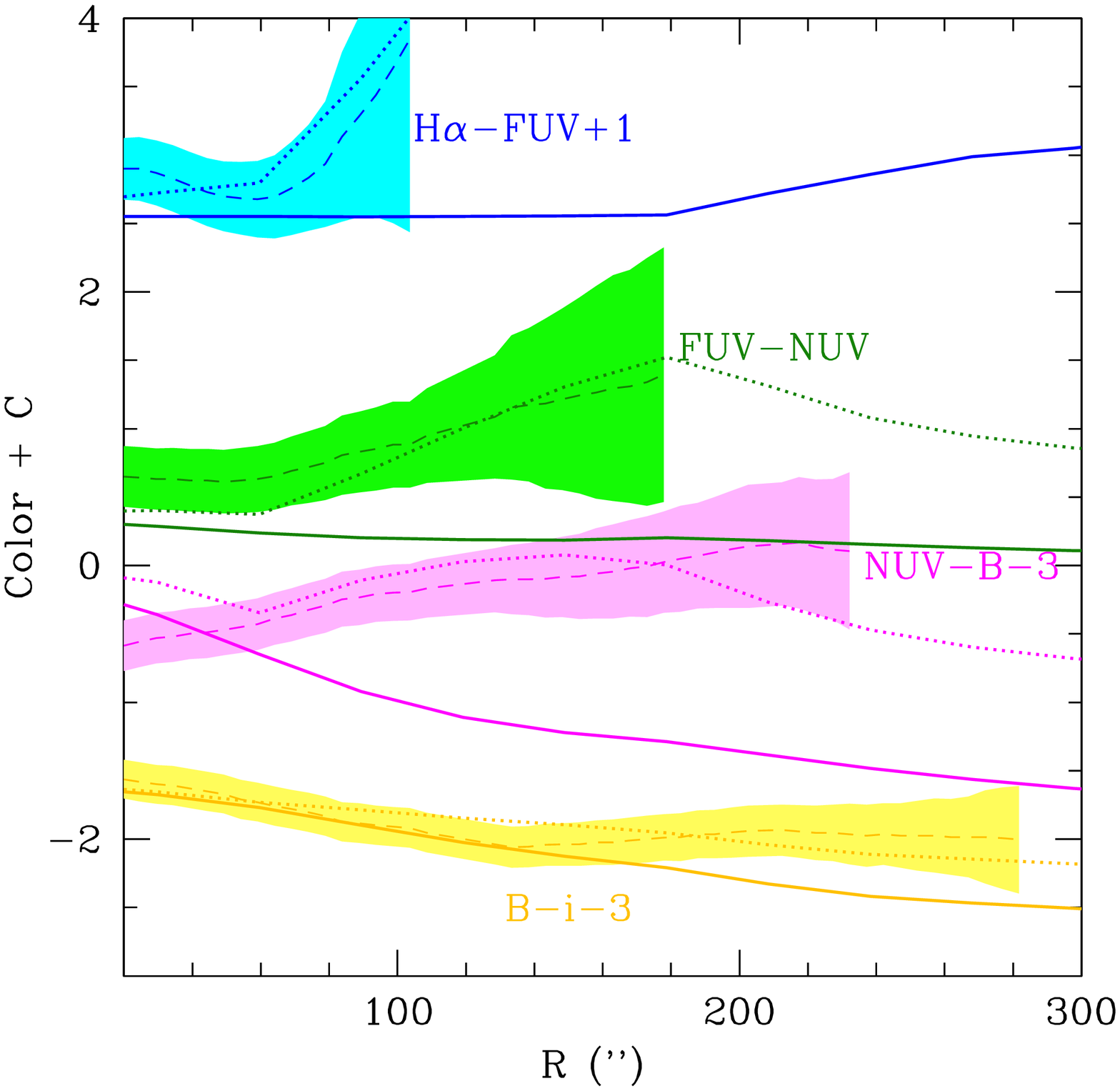}
%colorandsb.eps
\small{\caption{\label{figcolorsbprof}
Observed color profile (dashed curve) and uncertainties (shaded area).
The solid lines are the results of the unperturbed model, while the
dotted lines correspond to the best-model with ram-pressure.
%All the model profiles have been shifted
%so that the model with ram-pressure gives the best fit to the observations
%The blue dotted line gives the expected star formation profile
%as deduced from the total gas column density combined with the
%modified Schmidt law adopted in the present model.
%{\textbf PENSO CHE SIA PIÙ CHIARO FARE DEI PANNELLI DIVERSI PER OGNI BANDA. 
%E' IL CASO DI UNIFORMARE LE ALTRE FIGURE A QUESTA}
}}
\end{figure}
Among the uncertainties, the extinction corrections are certainly the
largest because the A(FUV) vs. TIR/FUV calibration is 
highly uncertain in such evolved galaxies where dust is not only heated by
UV photons but also by the general interstellar radiation field produced by more evolved stars.
Geometry and attenuation laws can vary from one galaxy to
another; and without a complete radiative transfer model, it is
impossible to apply ``perfect'' extinction corrections.  We present in
Figures \ref{FIGmodelREF} to \ref{figmodels} the data ``as observed''
(open symbols) and ``extinction corrected''(filled symbols) in order
to illustrate the sense and magnitude of the correction, and in doing
so, illustrate just how much the uncertainty affects our results.
It is obvious that allowing ourselves to change only two parameters ($t_{rp}$ and $\epsilon_0$)
and considering the constraint of 16 profiles (total gas+photometry+rotation curve)
does not result in  a perfect match for all of them, especially
taking into account the above-mentioned uncertainties.
% MAINTENANT DEJA DIT PLUS HAUT !
%We also remind that, as stated in the previous sections, extinction
%corrections can be at the origin of large uncertainties.
%Some noisy, small scale structures are still present in
%the smoothed profiles, observed mostly in the younger stellar
%populations (H$\alpha$, FUV and NUV). They cannot be reproduced by our
%smoothly evolving multi-zone model.
Models were computed each 100 Myr for 0 $<$ $t_{rp}$ $<$ 500 Myr,
200 Myr for 0.5 $<$ $t_{rp}$ $<$ 1.5 Gyr and 1 Gyr for 1.5 $<$ $t_{rp}$ $<$
6.5 Gyr; and in steps of 0.2 M$_{\odot}$ kpc$^{-2}$ yr$^{-1}$ efficiencies
between 0.4 and 1.6. Computing the $\chi^2$ for various values of
$\epsilon_0$ and $t_{rp}$ (see Fig.~\ref{figchi2}), we found that the
model best matching the profiles of NGC~4569 is characterized by
$\epsilon_0$ = 1.2 M$_{\odot}$ kpc$^{-2}$ yr$^{-1}$ and $t_{rp}$ = 100
Myr.
The $\chi^2$ for models with $t_{rp} <$ 400 Myr are still acceptable with $\chi^2$
lower than 5 times the $\chi^2$ of the best model ($\chi^2_{min}$=3.4).
%METTERE DELLE PERCENTUALI INVECE DEI VALORI DEL
%CHIQUADRO. 
The $\chi^2$ for models in disagreement with observational
limits at large radii (e.g. predicting too much gas or light in the
outer disk where none is observed) are artificially set to large
values to reject these solutions that cannot reproduce the observed
truncations (shaded area in Fig.~\ref{figchi2}).
Note that although the reduced $\chi^2$ was computed, with 2 free
parameters, the usual idea of a good fit ($\chi^2 \sim$ 1) cannot be
achieved since we know that our models will not reproduce the small
scale variations of the profiles (as mentioned above) and the
introduction of observational limits makes any statistical analysis
harder. In our study, the $\chi^2$ is only used to pick out the best
among the models, and see which parameters give similar results ($t_{rp} <$
400 Myr), and which parameters produce very unrealistic models (large
$t_{rp}$ and low $\epsilon_0$).
For a few profiles Fig.~\ref{figmodels} shows the observations contrasted with
models of various
ages ($t_{rp}$) and  of various efficiencies, showing that 
low and/or high  efficiencies do not reproduce properly the gas and H$\alpha$ truncation. 
In this figure, we also show a model with $t_{rp}$ = 300~Myr, the time indicated
by the dynamical model of Vollmer et al. (2004a).
It is interesting to note that although earlier cluster-core-crossing
epochs give more truncated disk profiles in the old stellar
populations (B and $i$ bands, blue dashed line), this is not the case
in the gas profile which is modified by contributions from the
recycled gas. Models with $t_{rp} >$ 500~Myr are quickly rejected
because recycled gas from evolving stars would be present
at large radii, thereby allowing some low levels of star formation.  
The model relative to the oldest interaction, $t_{rp}$=1.5 Gyr given 
in Fig.~\ref{figmodels} (dashed magenta line), predicts in fact a factor of $\sim$ two 
higher total gas surface density outside the 80 arcsec radius than that observed.

This is largely consistent with the dynamical models of Vollmer et
al$.$ (2004a), who obtained $t_{rp}$ $\sim$300~Myr, but do not reject
shorter ($\sim$100~Myr) timescales (Vollmer, private
communication). Although not reproducing perfectly the surface
brightness profiles, the best-fitting model is able to qualitatively reproduce
the truncation of the total gas disk profile and that of the H$\alpha$ and
UV radial profiles, as well as the milder truncation observed at longer
wavelengths (see Fig.~\ref{FIGmodelREF}). The comparison with the 
unperturbed model (assumed to represent an isolated spiral) 
clearly shows the effect of a ram-pressure-stripping event.

The ram-pressure model also reproduce (within the uncertainties) the
radial trends of colors which are also clearly different from the
unperturbed reference model. This is especially visible in
Fig.~\ref{figcolorsbprof} where we have compiled a few color profiles:
the observed data (smoothed to match the model resolution), the unperturbed
reference model, and the model including the perturbation. Color
gradients at short wavelengths are inverted (redder colors outward)
with respect to normal galaxies when  gas removal is introduced into
the model, as observed.

%\begin{figure*}
%\epsscale{1.0}
%\includegraphics[width=15cm,angle=0]{profileMODELrpSMOOTH.ps}
%\small{\caption{Profiles in NGC~4569. 
%Data are the same as in fig.~\ref{FIGmodelREF} The solid line is the
%model with a gas removal parameterized with $\epsilon_0$ = 1.2
%M$_{\odot}$ kpc$^{-2}$ yr$^{-1}$ and $t_{rp}$=100 Myr, the best model we
%adopted. Comparison with Fig.\ref{FIGmodelREF} (including the
%reference model) shows that our simple introduction of the gas removal
%in the model allows to reproduce the change in radial trends induced
%by the perturbation.
%{\textbf METTERE I PROFILI SMUSSATI.}
%\label{FIGmodelRP}}}
%\end{figure*}

%We acknowledge that our models are very simple (what allows us to 
%keep a minimum number of free parameter). 
The adopted model does not take into account possible effects
complicating the interaction. The most obvious one is the possibility
for some of the gas to be re-accreted onto the galaxy following the
stripping episode. This happens in several of the dynamical models run by Vollmer et
al. (2001b). The effect is strongly dependent on the inclination angle
of the galaxy with respect to its orbital plane, as well as the maximum ram
pressure. In all of their models having a significant inclination (i.e., larger
than 20 degrees), the re-accreted gas mass is lower than 10\% of the
stripped gas. If 10\% of the stripped gas in our model was
re-accreted (and distributed in a similar way to the gas profile we
obtained), the gas profile would increase only by 0.4 dex, which would
not be very different from the observed profiles. If we assume in the models
that a much larger fraction of the gas (about 50\%) is re-accreated, we
would obtain HI column densities exeeding the observed ones. Another reason why
we don't expect much re-accretion to have taken place is that this
phenomenon takes time: several 100 Myr years after the closest passage
of the galaxy to the cluster center, a time similar to the look-back
time we expect for this passage.

To conclude we can confidently say that the starvation scenario is very
unlikely to have shaped the NGC4569 profiles, while the ram-pressure
scenario is much more in agreement with the observations. With this
scenario, we can exclude  solutions older than $\sim$ 500~Myr, in
agreement with the dynamical models of Vollmer et al.
%Whatever perturbation is at the origin of the gas removal in NGC 4569, 
%the gas removal is certainly a relatively recent phenomenon.

\section{Discussion and conclusion}

\begin{figure}
\centering
\epsscale{1.0} \includegraphics[width=8cm,angle=0]{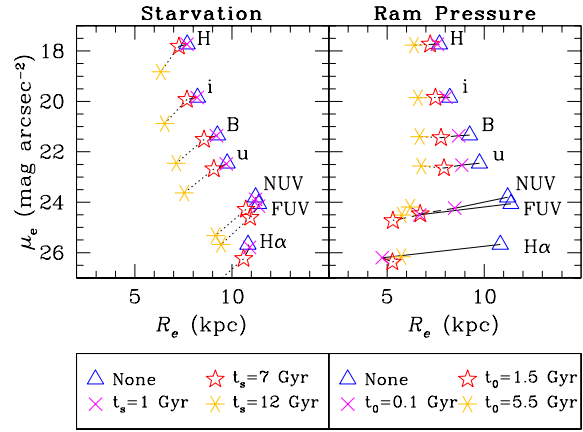}
\small{\caption{
Variation of the effective surface brightness
(mean surface brightness within $R_e$, the radius containing half of the total light)
and radius due to differential variation of the star formation history
of NGC~4569 in a) a starvation and b) ram-pressure stripping
scenario. Open triangles are for the unperturbed model, the other symbols 
for different ages of the interaction. }
\label{figmuchange}}
\end{figure}
\begin{figure*}
\epsscale{1.2}  
\plotone{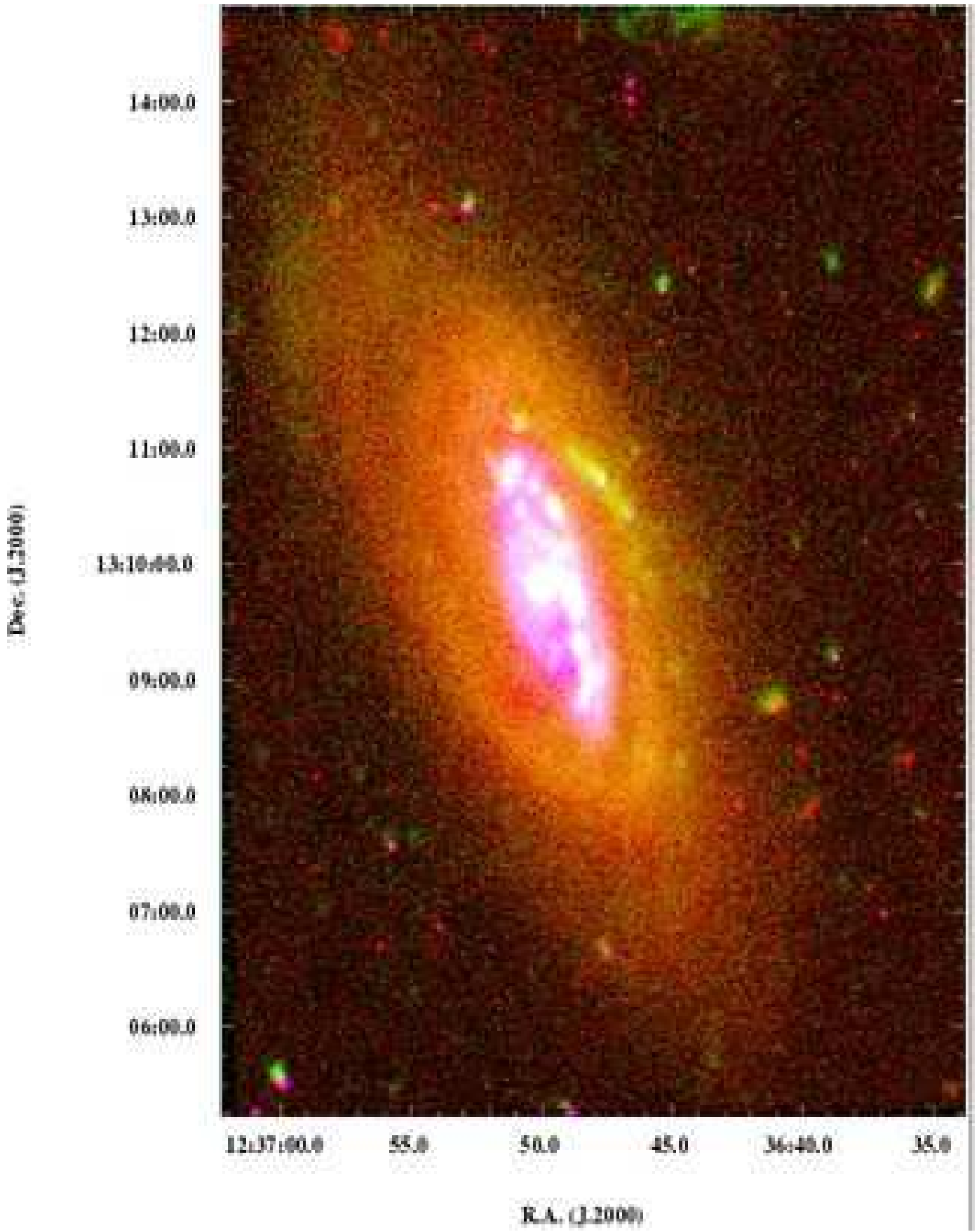} 
\small{\caption{The RGB (continuum
subtracted H$\alpha$ =blue, FUV=green, red continuum near H$\alpha$=red) color map of NGC~4569. The NW
spiral arm is visible at R.A. $\sim$ 12h37m47" and dec $\sim$ 13$^o$10'40".}
\label{RGB}
}
\end{figure*}

The present work gives the first quantitative estimate of the
structural evolution of stellar disks in cluster galaxies due to gas
removal caused by a dynamical interaction of the galaxy with the
IGM, delivering a strong message concerning the passive
stellar evolution of stripped disks. 
%Although the model only qualitatively reproduces the observed
%multi-wavelength radial profiles (the mismatch being attributed to
%resolution effects and large uncertainties in the extinction
%correction) it delivers a strong message concerning the passive
%stellar evolution of stripped disks. 
%NGC 4569 seems to be a representative
%example of HI-deficient, anemic cluster galaxies, with 
%, as it seems to be the
%case (Boselli \& Gavazzi 2006), 
%the results obtained from this analysis can be extended to a larger
%galaxy population. 
First of all it is clear that the truncation of the
total gas disk profile is soon reflected on the young population
stellar disk, confirming the predictions of Larson et al$.$ (1980). As
a consequence gas-stripped galaxies have color gradients opposite to
that of normal, isolated spirals, which generally have bluer colors in
their outer disks. NGC~4569 is bluest towards the center (see
{Fig. \ref{figcolorsbprof} and \ref{RGB}}). The trend is especially
true for colors tracing the relatively young populations ($< \sim$
10$^8$ yr); colors tracing populations older than the interaction
event show the usual gradient (i.e., redder towards the
center, as the B-$i$ color index). The inversion of the color gradient, here observed for the
first time in a cluster galaxy, is well reproduced by our model.\\ 
The second major conclusion of the present analysis is that the
perturbation which induced the truncation of the stellar disk of NGC 4569 is
relatively recent event ($\sim$100~Myr, in any case $\leq$ 500
Myr). Ram-pressure is favored with respect to starvation since the
latter is simply not able to reproduce the observed truncation of the 
gas and star forming (H$\alpha$, 8 $\mu$m, FUV) disk.
\\
Since NGC 4569 seems to be typical of the HI-deficient galaxy population
inhabiting nearby clusters, characterized by truncated HI and
star forming disks (Cayatte et al. 1994; Koopmann et al. 2006), with largely 
unperturbed older stellar populations\footnote{We remind that there is a 
strong relationship between the ratio of the optical to H$\alpha$ radii and 
the HI deficiency parameter in cluster galaxies (see Fig. 11 of Boselli \& Gavazzi 2006).}, we 
generalize this statement by saying that gas stripping in cluster galaxies is 
due to relatively recent perturbations.
This result is thus a major constraint on the evolution of cluster
galaxies since it rejects long time-scale phenomena such as 
harassment and starvation.\\
%
%The consequence of these findings in the interpretation of the
%evolution of cluster spiral galaxies is significant. The detailed
%analysis presented in Boselli \& Gavazzi (2006) has shown that the
%degree of truncation of the star forming disk in cluster spirals is
%proportional to their HI-deficiency parameter. At the same time the
%stellar disks observed in the R band are not truncated, as in NGC
%4569. This observational evidence allow us to conclude that NGC 4569
%is indeed representative of the HI-deficient galaxy population in
%nearby clusters and to generalize that the perturbation responsible
%for their gas depletion, whatever its nature is, is a relatively
%recent event ($\leq$ 500 Myr).  
%If gas stripping in
%earby clusters is primarily due to the interaction of galaxies with
%the cluster IGM (consistently with the recent work by Boselli \& Gavazzi 2006 and 
%with the specific case of NGC 4569 as indicated by this work or by Vollmer et al. 2004a) we can conclude
%that these galaxy-IGM interactions took place less than half a Gyr ago. 
%those phenomenon invoked to explain
%the decrease of the star formation activity (and HI gas deficiency)
%observed at large cluster-centric distances (harassment, starvation and
%pre-processing) just because their timescales are too long by a factor
%of $\sim$ 10.\\
The ram-pressure stripping scenario has recently been 
criticized as unable to reproduce the radial decrease with the
cluster-centric distance of the star formation activity of galaxies in
the nearby Universe.  Recent, complete spectroscopic surveys of the
nearby universe such as the SDSS or the 2dF have shown that the
activity of late-type galaxies begins to decrease at 1-2 virial radii
from the cluster center (Gomez et al. 2003; Lewis et al. 2002, Tanaka
et al. 2004, Nichol 2004), scale-lengths comparable to those observed
in the HI (Gavazzi et al. 2005, 2006a) and/or H$\alpha$ (Gavazzi et
al. 2002a, 2006b) in nearby clusters. Since these distances (1-2
virial radii) are significantly larger (a factor of 5-10) than those
where galaxy-IGM interactions are expected to be more efficient
($\sim$ inside a core radius), other processes already active at the
periphery of clusters [such as galaxy harassment (Moore et al. 1996),
starvation (Balogh et al. 2000) or pre-processing (Fujita 2004; Cortese et al. 2006b)] have
been proposed to explain the radial decrease of the star formation
activity with the cluster-centric distance.\\ 
As extensively discussed
in Boselli \& Gavazzi (2006), however, the presence of galaxies with
clear signs of ongoing ram-pressure stripping at the periphery of
nearby clusters and the large velocity dispersion, combined with the
radial orbits (Dressler 1986; Solanes et al. 2001) of the late-type
galaxy population can explain the observed decrease of the star
forming activity at large cluster-centric radii in a ram-pressure
stripping scenario.  NGC 4569, for instance, might have travelled
$\sim$0.6 Mpc (1/3 of the virial radius) since it was stripped, a
distance significantly larger than the core radius of the cluster (130
kpc). We recall that in Virgo the drop of the star formation in bright
galaxies is observed at $\sim$0.7 virial radii (Gavazzi et al. 2006b),
while the increase of the HI-deficiency parameter occurs at about 1 virial
radius (Gavazzi et al. 2005). The higher velocity dispersion and IGM
gas density of rich, evolved clusters, as well as a clumpy IGM
distribution, might thus be at the origin of ram-pressure stripped
galaxies up to $\sim$ one virial radius. \\

One of the most intriguing (and still open)
questions regarding the effects of the environment on the evolution of
galaxies is that of the origin of lenticulars, and their overabundance
in the centers of rich clusters. Are lenticulars an independent
population of galaxies formed in the primordial high-density
environments, or were they spiral disks whose star formation activity
had been quenched once their gas reservoir was removed by the
unfavorable cluster environment?
%Answering this question unequivocally is a challenge. 
The present work has shown for the first time how a
galaxy-cluster IGM interaction is able to remove most of the gas
reservoir, inducing important structural modifications in the
stellar disk properties. We have in fact shown that, because of the
differential radial stellar evolution of spiral disks, we can expect that
cluster spirals have (at least at short wavelengths) more truncated disk
profiles, inverting the outer color gradient with respect to similar
but unperturbed objects. The surface
brightness of the disk, however, mildly decreases in H$\alpha$ and in
the UV bands while remaining mostly constant at longer wavelengths,
even 5~Gyr after the interaction
(Fig.~\ref{figmuchange}) \footnote{40\% of  S0 galaxies in the Coma cluster underwent
a star formation event in their centers in the last 5 Gyr, Poggianti et al. (2001).}.  
The differential evolution of the stellar disk due to gas
stripping alone is thus not able to reproduce the $\sim$ 0.65 mag increased
central surface brightness of present-day lenticulars (Dressler 1980; Boselli \& Gavazzi
2006). In the case of starvation stellar disks are not truncated ($R_e$ just slightly decreases) 
while surface brightnesses significantly decrease on long time scales.\\
Gravitational perturbations, such as tidal
interactions with other galaxies (Merritt 1983), interactions with the
cluster potential well (Byrd \& Valtonen 1990) or a mixture of both
%(called `galaxy harassment' by Moore et al$.$ 1996)
must be invoked to reproduce the observed properties of nearby
lenticulars.  As described in the introduction, similar conclusions
have been obtained from the analysis of statistical, kinematical,
structural and spectro-photometrical properties of nearby clusters.\\
This new study is consistent with the idea that the present evolution
of late-type galaxies in clusters differs from that at earlier epochs,
where late-type galaxies were mostly perturbed by dynamical
interactions (pre-processing and/or galaxy harassment; Dressler 2004,
Moore et al$.$ 1996) which were able to thicken the stellar disks
thereby producing the present-day cluster lenticulars. \\ To conclude
we can say that the combination of multifrequency observations of
spatially resolved cluster galaxies combined with multi-zone
spectro-photometric models of galaxy evolution provides an extremely useful
tool to study the evolution of cluster galaxies. The results here
presented for a typical HI-deficient, anemic Virgo cluster galaxy and
extrapolated to the whole cluster galaxy population indicate that:\\
1) The gas removal due to a ram-pressure stripping event can reproduce
the observed radial truncation of the stellar disk stronger at shorter
wavelengths. On the other hand,  starvation is not able to truncate gas or
stellar disks.\\ 
2) As a consequence, color gradients make cluster
objects redder in the outer disk than in the inner regions, and are thus
inverted with respect to normal, isolated late-type galaxies.\\ 
3) This technique is useful to age-date the interaction. If NGC 4569 
represents the 
typical HI-deficient cluster galaxy with truncated H$\alpha$ and HI
disk and normal intermediate age stellar disk, then our modeling suggests that gas
stripping is a relatively recent event since it probably took place $\leq$500 Myr ago in these systems.\\ 
4) The effective surface brightness of the stripped galaxies 
remains constant or even mildly
decreases after the interaction.\\ 
We hope to confirm this original
result in the near future once multi-frequency data come available for
a statistical significant sample of late-type cluster galaxies.

\acknowledgements GALEX (Galaxy Evolution Explorer) is a NASA Small
Explorer, launched in April 2003.  We gratefully acknowledge NASA's
support for construction, operation, and science analysis for the
GALEX mission, developed in cooperation with the Centre National
d'Etudes Spatiales of France and the Korean Ministry of Science and
Technology.  We thank G. Gavazzi and N. Prantzos for their long-term
collaboration in the subjects studied in this paper. S.B. thanks the
CNES for its funding through GALEX-Marseille. We wish to thank
B. Vollmer and J. Kenney for their valuable comments, and the GALEX
SODA team for their help in the data reduction.

\references

\reference{}Abazajian, K., et al$.$, 2005, 129, 1755

\reference{}Balogh, M.L., Navarro, J.F., \& Morris, S.L., 2000, ApJ, 540, 113

\reference{}Boissier, S., \&  Prantzos, N.\ 1999, MNRAS, 307, 857 

\reference{}Boissier, S. \& Prantzos, N., 2000, MNRAS, 312, 398 

\reference{}Boissier, S., Boselli, A., Prantzos, N., \& Gavazzi, G.\ 2001, MNRAS, 321, 733 

\reference{}Boissier, S., Prantzos, N., Boselli, A. \& Gavazzi, G., 2003, MNRAS, 346, 1215 

\reference{}Boissier, S., Boselli, A., Buat, V., Donas, J. \& Milliard, B., 2004, A\&A, 424, 465

\reference{}Boselli, A. \& Gavazzi, G., 2002, A\&A, 386, 124

\reference{}Boselli, A. \& Gavazzi, G., 2006, PASP, 118, 517

\reference{}Boselli, A., Tuffs, R., Gavazzi, G., Hippelein, H. \& Pierini, D., 1997, A\&AS, 121, 507

\reference{}Boselli, A., Gavazzi, G., Donas, J. \& Scodeggio, M., 2001, AJ, 121, 753

\reference{}Boselli, A., Lequeux, J. \& Gavazzi, G., 2002, A\&A, 384, 33

\reference{}Boselli, A., Gavazzi, G. \& Sanvito, G., 2003, A\&A, 402, 37

\reference{}Boselli, A., Lequeux, J., Gavazzi, G., 2004, A\&A, 428, 409

\reference{}Buat, V., Iglesias-Paramo, J., Seibert, M., et al., 2005, ApJ, 619, L51

\reference{}Byrd, G. \& Valtonen, M., 1990, ApJ, 350, 89

\reference{}Cayatte, V., van Gorkom, J., Balkowski, C., \& Kotanyi, C., 1990, AJ, 100, 604

\reference{}Cayatte, V., Kotanyi, C., Balkowski, C.\& van Gorkom, J., 1994, AJ, 107, 1003

\reference{}Christlein, D., Zabuldoff, A.I., 2004, ApJ, 616, 192 

\reference{}Cortese, L., Boselli, A., Buat, V., et al., 2006a, ApJ, 637, 242

\reference{}Cortese, L., Gavazzi, G., Boselli, A., Franzetti, P., Kennicutt, R., O'Neil, K., Sakai, S., 2006b, A\&A, in press
(astro-ph/0603826)

\reference{}Cowie, L.L., \& Songaila, A., 1977, Nat., 266, 501

\reference{}Dale, D., Helou, G., Contursi, A., Silbermann, N., Kolhatkar, S., 2001, ApJ, 549, 215

\reference{}Dale, D., Bendo, G., Engelbracht, C., et al., 2005, ApJ, 633, 857

\reference{}Dressler, A., 1980, ApJ, 236, 351

\reference{}Dressler, A., 1986, ApJ, 301, 85

\reference{}Dressler, A., in "Clusters of Galaxies: Probes of Cosmological Structure 
and Galaxy Evolution", Cambridge University Press, ed. by Mulchaey et al., 2004, p. 207

\reference{}Fujita, Y., 1998, ApJ, 509, 587

\reference{}Fujita, Y., 2004, PASJ, 56, 29

\reference{}Fujita, Y., Nagashima, M., 1999, ApJ, 516, 619

\reference{}Gavazzi, G., Pierini, D. \& Boselli, A., 1996, A\&A, 312, 397  

\reference{}Gavazzi, G., Franzetti, P., Scodeggio, M., Boselli, A. \& Pierini, D., 2000, A\&A, 361, 863

\reference{}Gavazzi, G., Boselli, A., Mayer, L., Iglesias-Paramo, J., Vilchez, J.M., Carrasco, L., 2001, ApJ, 563, L23

\reference{}Gavazzi, G., Boselli, A., Pedotti, P., Gallazzi, A. \& Carrasco, L., 2002a, A\&A, 396, 449

\reference{}Gavazzi, G., Bonfanti, C., Sanvito, G., Boselli, A., \& Scodeggio, M., 2002b, ApJ, 576, 135 

\reference{}Gavazzi, G., Boselli, A., Donati, A., Franzetti, P. \& Scodeggio, M., 2003, A\&A, 400, 451

\reference{}Gavazzi, G., Zaccardo, A., Sanvito, G., Boselli, A. \& Bonfanti, C., 2004, A\&A, 417, 499

\reference{}Gavazzi, G., Boselli, A., van Driel, W., O'Neil, K., 2005, A\&A, 429, 439

\reference{}Gavazzi, G., O'Neil, K., Boselli, A., van Driel, W., 2006a, A\&A, 449, 929

\reference{}Gavazzi, G., Boselli, A., Cortese, L., Arosio, I., Gallazzi, A., Pedotti, P., Carrasco, L., 2006b, A\&A, 443, 839

\reference{}Gil de Paz, A. \& Madore, B., 2005, ApJS, 156, 345

\reference{}Gil de Paz, A., Boissier, S., Madore, B., et al., 2006, ApJS, in press 

\reference{}Gomez, P.L., Nichol, R.C., Miller, C.J., et al., 2003, ApJ, 584, 210

\reference{}Gunn, J.E., \& Gott, J.R.I., 1972, ApJ, 176, 1

\reference{}Jarrett, T., Chester, T., Cutri, R., Schneider, S. \& Huchra, J., 2003, AJ, 125, 525

\reference{}Jogee, S., Scoville, N., Kenney, J., 2005, ApJ, 630, 837

\reference{}Helfer, T., Thornley, M., Regan, M., et al, 2003, ApJS, 145, 259

\reference{}Hinz, J.L., Rieke, G.H., \& Caldwell, N., 2003, AJ, 126, 2622

\reference{}Hou, J.~L., Prantzos, N., \& Boissier, S.\ 2000, A\&A, 362, 921 

\reference{}Iglesias-Paramo, J., Boselli, A., Gavazzi, G., Zaccardo, A., 2004, A\&A, 421, 887

\reference{}Kenney, J. \& Young, J., 1988, ApJS, 66, 261

\reference{}Kenney, J., van Gorkom, J. \& Vollmer, B., 2004, ApJ, AJ, 127, 3361

\reference{}Kennicutt, R., Armus, L., Bendo, G., et al., 2003, PASP, 115, 928

%\reference{}Kennicutt, R., 1998, ARA\&A, 36, 189

\reference{}Koopmann, R. \& Kenney, J., 2004a, ApJ, 613, 851

\reference{}Koopmann, R. \& Kenney, J., 2004b, ApJ, 613, 866

\reference{}Koopmann, R., Haynes, M., Catinella, B., 2006, AJ, 131, 716

\reference{}Kroupa, P., Tout, C.~A., \& Gilmore, G.\ 1993, MNRAS, 262, 545 

\reference{}Larson, R.~B.\ 1976, MNRAS, 176, 31 

\reference{}Larson, R., Tinsley, B. \& Caldwell, N., 1980, ApJ, 237, 692

\reference{}Lewis, I., Balogh, M., De Prorpis, R., et al., 2002, MNRAS, 334, 673

\reference{} Martin, C.~L., \&  Kennicutt, R.~C.\ 2001, \apj, 555, 301 

\reference{}Merritt, D., 1983, ApJ, 264, 24

\reference{}Mo, H.~J., Mao, S., \&  White, S.~D.~M.\ 1998, MNRAS, 295, 319 

\reference{}Moore, B., Katz, N., Lake, G., Dressler, A. \& Oemler, A., 1996, Nat., 379, 613

\reference{}Nichol, R.C., 2004, in Clusters of Galaxies: Probes of Cosmological Structure and Galaxy Evolution, p24

\reference{}Nulsen, P.E.J., 1982, MNRAS, 198, 1007

\reference{}Poggianti, B.M., Bridges, T.J., Carter, D., et al., 2001, ApJ, 563, 118

\reference{}Poggianti, B.M., Bridges, T.J., Komiyama, Y., et al., 2004, ApJ, 601, 197 

\reference{}Prantzos, N., \& Boissier, S.\ 2000, MNRAS, 313, 338 

\reference{}Rubin, V., Waterman, A. \& Kenney, J., 1999, AJ, 118, 236

\reference{}Samland, M., Hensler, G., \& Theis, C.\ 1997, ApJ, 476, 544 

\reference{}Solanes, J.M., Marrique, A., Garcia-Gomez, C., Gonzales-Casado, G., Giovanelli, R., Haynes, M., 2001, ApJ, 548, 97

\reference{}Tanaka, M., Goto, T., Okamura, S., Shimasaku, K., \& Brinkmann, J., 2004, AJ, 128, 2677

\reference{}Tinsley, B.~M.\ 1980,  Fundamentals of Cosmic Physics, 5, 287 

\reference{}Treu, T., Ellis, R.~S., 
Kneib, J.-P., Dressler, A., Smail, I., Czoske, O., Oemler, A., \& 
Natarajan, P.\ 2003, ApJ, 591, 53 
 
\reference{}van den Bergh, S., 1976, ApJ, 206, 883

\reference{}V{\'a}zquez, G.~A., \& Leitherer, C.\ 2005, ApJ, 621, 695 
 
\reference{}Vollmer, B., Cayatte, V., Boselli, A., Balkowski, C. \& Duschl, W., 1999, A\&A, 349, 411

\reference{}Vollmer, B., Marcelin M., Amram, P., Balkowski, C., Cayatte, Garrido, O., 2000, A\&A, 365, 532

\reference{}Vollmer, B., Braine, J.,  Balkowski, C., Cayatte, V. \& Duschl, W., 2001a, A\&A, 374, 824

\reference{}Vollmer, B., Cayatte, V., Balkowski, C. \& Duschl, W., 2001b, ApJ, 561, 708

%\reference{}Vollmer, B., 2003, A\&A, 398, 525

\reference{}Vollmer, B., Balkowski, C., Cayatte, V., van Driel, W. \& Huchtmeier, W., 2004a, A\&A, 419, 35 

%\reference{}Vollmer, B., Huchtmeier, W., van Driel, W., 2004b, A\&A, 439, 921

\reference{}Vollmer, B., Beck, R., Kenney, J., van Gorkom, J., 2004b, AJ, 127, 3375 

\reference{}Vollmer, B., Braine, J., Combes, F., Sofue, Y., 2005, A\&A, 441, 473

\reference{}Whitmore, B.C., Gilmore, D.M., \& Jones, C., 1993, ApJ, 407, 489 

\reference{}Yoshida, M., Ohyama, Y., Iye, M., et al., 2004, AJ, 127, 90

\end{document}